\documentclass[%
 reprint,
 amsmath,amssymb,
 aps,
]{revtex4-2}

\usepackage[english]{babel}

\usepackage[a4paper,top=2cm,bottom=2cm,left=3cm,right=3cm,marginparwidth=1.75cm]{geometry}

\usepackage{amsmath}
\usepackage{graphicx}
\usepackage{dcolumn}
\usepackage{bm}
\usepackage[colorlinks=true, allcolors=blue]{hyperref}

\begin{document}

\title{Towards Baikal-Top: Feasibility study of an onshore detector system for the joint registration of EAS with Baikal-GVD}

\author{E.A.~Kravchenko}%
 \email{E.Kravchenko@nsu.ru}
\affiliation{%
 Budker Institute of Nuclear Physics SB RAS,\\
pr.Lavrentieva 11, Novosibirsk, Russia}
\affiliation{%
Novosibirsk State University,\\
Pirogova str. 2, Novosibirsk, Russia}
\affiliation{%
Institute for Nuclear Research of the Russian Academy of Sciences,\\
60th October Anniversary st. 7A, Moscow, Russia
}%

\author{G.I.~Rubtsov}
\affiliation{%
Institute for Nuclear Research of the Russian Academy of Sciences,\\
60th October Anniversary st. 7A, Moscow, Russia
}%

\author{D.S.~Zhadan}
\affiliation{%
Budker Institute of Nuclear Physics SB RAS,\\
pr.Lavrentieva 11, Novosibirsk, Russia}
\affiliation{%
Novosibirsk State University,\\
Pirogova str. 2, Novosibirsk, Russia
}%


\date{\today}

\begin{abstract}
We study the possibility of registering high-energy extensive air showers (EAS) by the onshore detector facility simultaneously with the trigger of the Baikal-GVD neutrino telescope. The location of the surface detector array on the shore of Lake Baikal is motivated by the fact that permanent placement of detectors on the surface of the lake is challenging. Within the given geometry, simultaneous registration is possible for EAS with a zenith angle of about 76 degrees within the limited solid angle. The corresponding inclined EAS are dominated by muons and significantly attenuated. The installation will make it possible to obtain an estimate of the number of high-energy muons in EAS. This, subsequently, would make it possible to verify EAS modeling and calculations of the atmospheric neutrino flux. The detector may also be used for cross-calibration of energy and direction measurements by the neutrino telescope. We use the CORSIKA program to simulate the registration of EAS on the shore close to the Baikal-GVD. The propagation of ultra-high energy muons produced by EAS through 3.5~km of water and their registration by Baikal-GVD is simulated using the PROPOSAL package. We calculate the minimal total area of the onshore detectors suitable for several cosmic ray energy thresholds, starting from 1 PeV. The report presents estimates of the number of events jointly registered by the onshore installation and the Baikal-GVD the EAS registration detector systems with different areas.
\end{abstract}

\maketitle

\section{Introduction}
There are currently three operational neutrino telescopes that detect high-energy neutrinos ($>$ 1 TeV): the IceCube neutrino observatory (IceCube)~\cite{IceCube} constructed under the Antarctic ice at the South Pole, the Baikal Deep Underwater Neutrino Telescope (Baikal-GVD)~\cite{Avrorin:2020dre} in the water of Lake Baikal and the Cubic Kilometre Neutrino Telescope (KM3NeT)~\cite{KM3Net} at the bottom of the Mediterranean Sea.

The primary scientific objective of neutrino telescopes is to detect high-energy neutrinos from astrophysical sources. Furthermore, an important task for particle astrophysics is the detection of cosmogenic neutrinos produced by the interaction of ultra-high-energy protons and nuclei with radiation during their propagation through the intergalactic medium. Reliable detection of neutrinos of extraterrestrial origin requires an accurate accounting of background events. The primary source of background is atmospheric neutrinos produced in the decay of muons and pions in EAS caused by high-energy cosmic particles. By measuring the flux of high-energy atmospheric muons, it is possible to verify the flux of atmospheric neutrinos of comparable energies.

The IceCube observatory includes a detector array on the ice surface called IceTop. IceTop acts as a veto detector~\cite{IceCube:2025ary} for some downgoing events. Detecting muons in IceCube simultaneously with EAS on IceTop made it possible to estimate the number of high-energy muons in the EAS~\cite{IceCube:2025baz}. Furthermore, the IceTop experiment measured the spectrum and composition of cosmic rays in the energy range from PeV to EeV~\cite{IceCube:2019hmk}. Simultaneous measurement of high- and low-energy muons in the same shower has been shown to be important for testing high energy hadronic interactions and for solving the long-standing muon puzzle~\cite{Pyras:2025rjf}.

An expansion of the IceTop experiment is currently planned, which will include scintillation panels and radio antennas in addition to the ice tanks, with the goal of extending the aperture for coincident IceCube and IceTop events by a factor of 30~\cite{IceCube:2023uyr}. Moreover, an array of imaging air Cherenkov telescopes (IceAct) is being deployed on the surface above the IceCube~\cite{IceCube:2025iss}. The flux of high energy muons has recently been measured by the KM3NeT underwater detector~\cite{KM3NeT:2024buf} and by the China Jinping Underground Laboratory~\cite{Zhang:2025msm}. 

This paper proposes the design of an onshore detector capable of detecting extensive air showers (EAS) simultaneously with the trigger of the Baikal-GVD neutrino telescope. We have performed modeling of EAS in the atmosphere and subsequent particle propagation in water. An estimate is obtained for the number of joint events between the ground-based (onshore) detector and the Baikal-GVD. It is shown that joint detection allows to measure the high-energy muon flux in EAS. However, using the ground-based facility as a veto detector for a neutrino telescope would require detector areas that exceed the capabilities of the potential site. The detector can be used to validate atmospheric neutrino background modeling and for energy and direction cross-calibration.

\section{Simulation model}

\begin{figure*}
\centering
\includegraphics[width=0.9\linewidth]{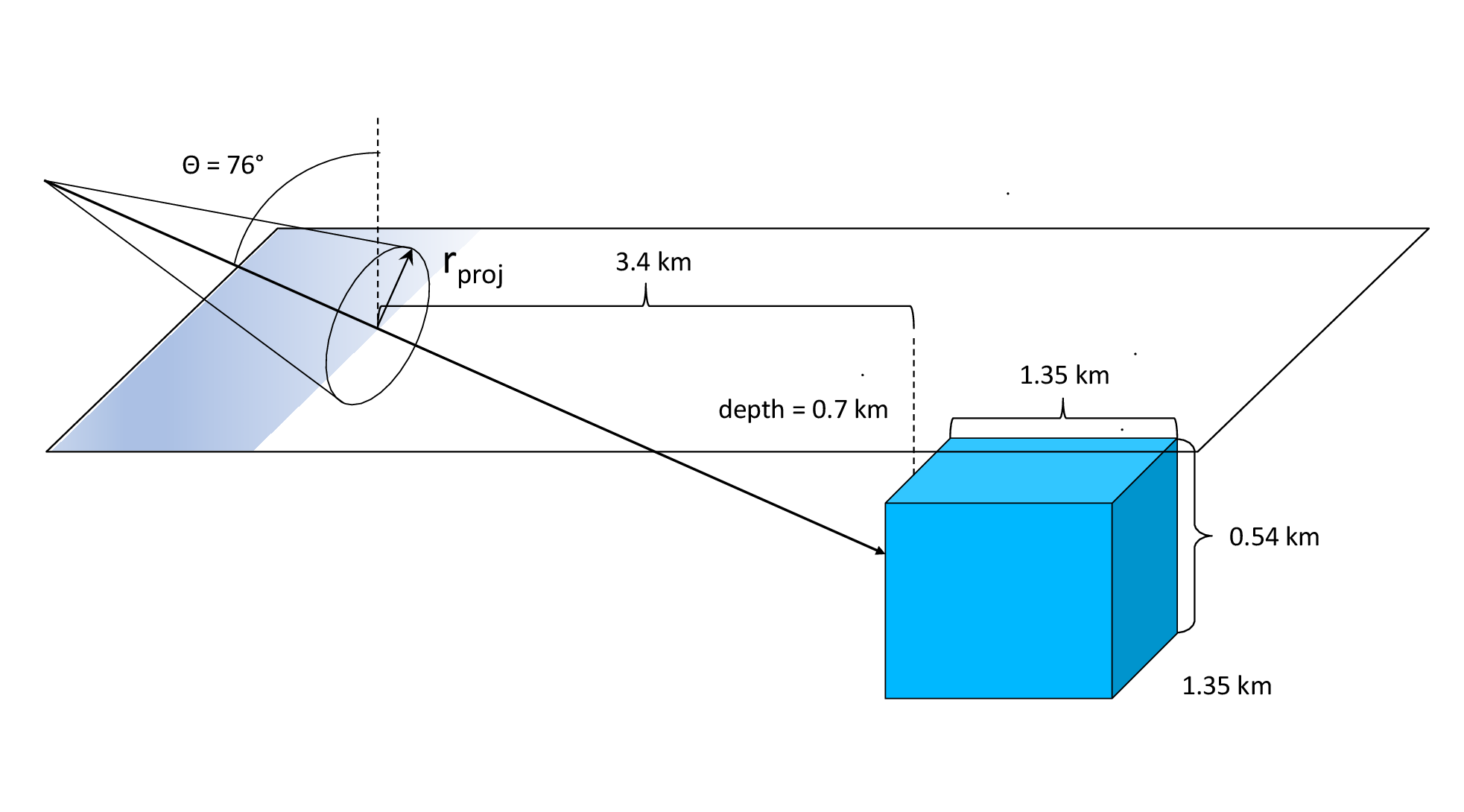}
\caption{\label{fig:SchemeSim}Scheme of the simulation model.}
\end{figure*}

Scheme of the simulation model presented in Figure~\ref{fig:SchemeSim}. To simulate cosmic particles, we use the CORSIKA\cite{corsika} program (Version 7.7550). In our work, we configured it with the following options: for the high-energy hadron interaction model, QGSJETII-04 was used; for the low-energy hadron interaction, GHEISHA 2002d was used; the horizontal flat detector array was selected for the detector geometry, and from the additional list of CORSIKA program options, CURVED and NEUTRINO options were chosen. The CURVED option is specific to our task since, at the incidence angle of 76 degrees, the geometry and thickness of the atmosphere begin to contribute. Moreover, without this option, modeling at such an angle in CORSIKA is not possible. Besides that, the program needs a configuration file for running, where parameters such as energy, angles, the type of primary particle, etc. are set. Here we set the parameters of interest for our task, for example, the observation level at Lake Baikal -- 452~m and energy cuts for secondary particles -- 0.3~GeV for hadrons and muons, 0.003~GeV for electrons and photons. In the simulation output, we receive binary files, which we then process with a special CorsikaPlotter utility from the Coast package (included in the CORSIKA program) to obtain files in ROOT\cite{Root} format.

There are several methods of working with ROOT files, we have chosen a simple one, Python and pyROOT\cite{pyROOT}. We have implemented a module on Python and used it in our analysis. This module includes the following functionalities: obtaining the number of certain particles, retrieving the energy distribution, and obtaining the integral and differential density distributions at a given radius. The latter may be either the actual radius on the ground or the core distance, the projection of a radius onto the plane perpendicular to the shower axis.

\begin{figure}
\centering
\includegraphics[width=1.0\linewidth]{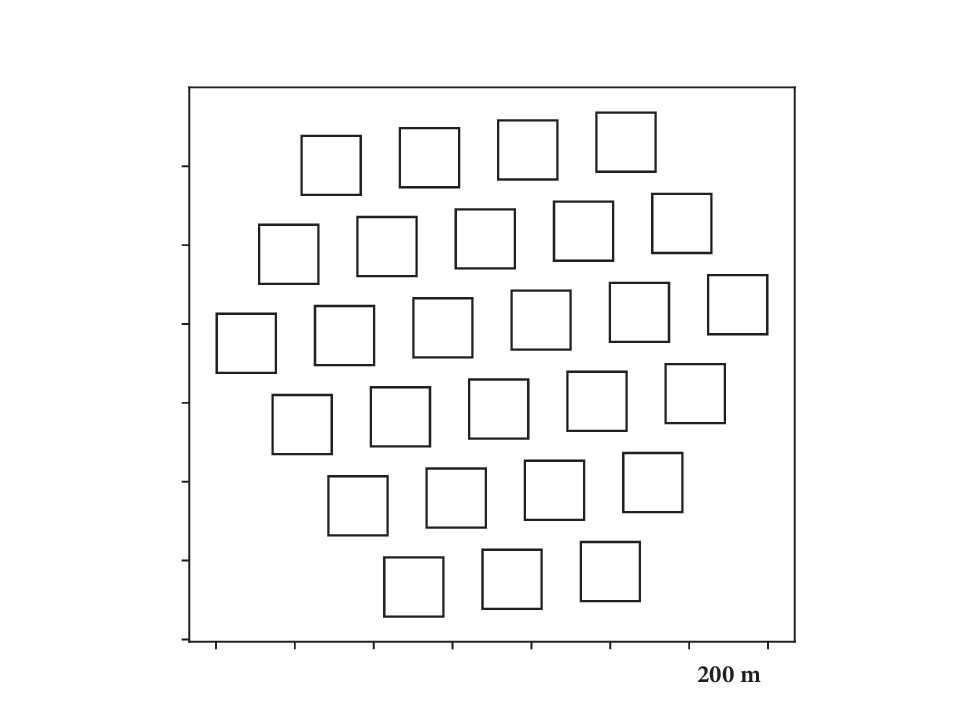}
\caption{\label{fig:DetCl} Description of Baikal-GVD clusters in the simulation model, top view.}
\end{figure}

The above applies to EAS development up to the surface of the water; we are also interested in what occurs in the water column on the way to the Baikal-GVD detector. For this purpose, we use the Proposal package~\cite{Koehne:2013gpa}. Proposal uses special classes named Propagators, which provide particle simulation through the water. For different types of particles, there are different Propagators. We use two types: mu minus and mu plus Propagators. In addition, a description of the geometry and material configuration is provided.  In our case, everything is water, with a special location of the detector clusters, where the energy cut that separates continuous and stochastic losses is set to a lower value. Within the special region, losses are considered stochastic if the energy transfer is greater than 500 MeV, whereas in the remaining volume of water the losses are considered stochastic if the relative energy transfer is greater than $5\%$. The Baikal-GVD clusters were described as vertical square columns with a shape of $150\times150\times540$~m, the position of the columns matches the description of the final configuration of the Baikal-GVD (Fig.~\ref{fig:DetCl}) \cite{Baikal-GVDSTR2012}. It should be mentioned that we simulate EAS with one fixed geometry, and hence our estimate does not account for the effects near the boundary of Baikal-GVD.

\section{Peculiarities of inclined EAS registration}

The location of Baikal-GVD, its depth, and distance from the shore determine the characteristic zenith angle of the EAS at which it can be detected simultaneously on the surface of the lake shore and in the depths of the water. Assuming that the average depth of the Baikal-GVD modules is 1~km and that the distance from the shore to the center of the neutrino telescope is 4 km, we obtain a characteristic angle equal to 76 degrees.

The registration of highly inclined EAS has several peculiarities. Unlike vertical EAS, the main contribution to the number of charged particles at ground level is made by muons, not electrons. In addition, due to the long length of the shower, the characteristic transverse size increases significantly. For example, 50\% of the muons are concentrated in a circle with a radius of about 1 km at a zenith angle $76^\circ$ degrees.
In the Table~\ref{tab:EASdata} the dependencies of the number of secondary particles in an EAS on the zenith angle, as well as the characteristic radii within which 50\% of the particles are contained are presented for an EAS initiated by a proton with an energy of 1~PeV.

\begin{table*}
\centering
\caption{\label{tab:EASdata} EAS generated by a 1 PeV proton. The dependencies of the number of secondary particles in an EAS on the zenith angle and the characteristic radii within which 50\% of the particles are contained.}
\begin{tabular}{l|c|c|c|c|c|c}
\hline
                        & 0$^\circ$ & 20$^\circ$ & 40$^\circ$   & 60$^\circ$    &{\bf 76}$^\circ$   &80$^\circ$ \\\hline
muons                   & 10546     & 10156     & 7973          & 4628          & {\bf1900 }        & 1312      \\
electrons(positrons)    & 132236    & 97631     & 20221         & 1123          & {\bf474  }        & 351       \\
gamma-quanta            & 696608    & 524195    & 115280        & 5315          & {\bf2076 }        & 1534      \\
neutrinos                & 22722     & 24209     & 26860         & 28492         & {\bf25858 }       & 23383     \\
R$_{pr}(\mu)$, m, (50\%)    & 256   & 275       & 352           & 522           & {\bf1221  }       & 1943      \\
R$_{pr}(e^+e^-)$, m, (50\%) & 32    & 34        & 49            & 413           & {\bf928  }        & 1336      \\
R$_{pr}(\gamma)$, m, (50\%) & 63    & 67        & 94            & 441           & {\bf943  }        & 1358      \\
R$_{pr}(\nu)$, m, (50\%)    & 600   & 682       & 1016          & 1902          & {\bf4733 }        & 6682      \\\hline
\end{tabular}
\end{table*}

The simplest way to detect EAS is to have a detection system of minimal ionizing particles of a rather large area, consisting of several sub-detectors between which it is possible to organize a system of signal coincidence for the trigger.
The count rate of EAS events will depend on the requirements for the minimum number of secondary particles detected in an event and the area of the detection system.

To estimate the frequency of events, we fixed the maximum distance (radius) from the detection system to the EAS axis at 100 m, at which reliable shower registration is expected.  We assume that the shower is registered efficiently if expected number of secondary particles hitting detector is 3 or larger. Knowing the density of secondary charged particles $f(r)$ at a distance of 100~m from the shower axis, the required particle detection area can be estimated: $S = 3/f$(100~m).

\begin{figure*}
\centering
\includegraphics[width=0.8\linewidth]{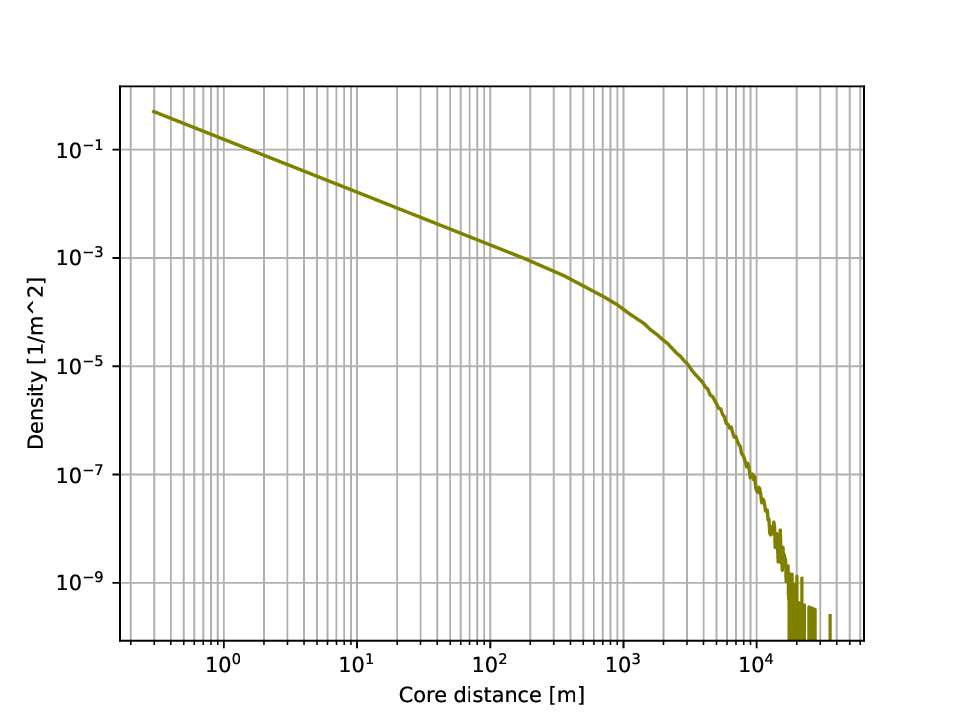}
\caption{\label{fig:Mu+eDens}Particle density function for muons and electrons (+positrons) greater than 10~MeV for the EAS generated by 1~PeV proton. Calculations are presented for the inclined plane (perpendicular to EAS axis).}
\end{figure*}

The particle density function $f(r)$ for secondary muons and electrons (+positrons) with energy greater than 10~MeV for the EAS generated by 1~PeV proton is presented in Figure~\ref{fig:Mu+eDens}. For 1~PeV shower $f$(100~m) = 0.0018~[part./m$^2$]. This gives us an estimate that with the detection system of $\sim$1700 m$^2$ it is possible to register 1 PeV (and higher) inclined at 76$^\circ$ EAS in a radius circle of 100 m.

The data on integral primary cosmic particle flux (J) is known, for example, \cite{Grieder2010}. At the 1~PeV threshold  $J=3\cdot10^{-6}$ [part./(m$^2$$\cdot$s$\cdot$sr)]. The solid angle of the EAS registration is defined by the height (0.54~km) and width (1.35~km) of the Baikal-GVD detection system and the mean distance to the shore (4~km) -- $\Omega = 0.54\times1.35/4^2=0.046$\,[sr]. The effective area of registration for EAS is $\pi\times(100\,\mbox{m})^2=3.1\times 10^4$\,m$^2$. Taking everything together, we estimate the event rate of the detection system of 1700\,m$^2$ area at the level of $1.4\cdot10^{5}$ events in a year with a 1~PeV threshold. Analogous estimations have been made for the EAS detection thresholds 3, 10 and 30~PeV, and the characteristic zenith angle 76$^\circ$  (Table~\ref{tab:EASrate}).

\begin{table*}
\caption{\label{tab:EASrate} Particle densities at radius of 100~m (inclined plane), estimated area of registration, integral primary cosmic particle flux at given energy threshold, event rate estimations.}
\centering
\begin{tabular}{l|c|c|c|c} \hline
E threshold                         & 1 PeV             & 3 PeV             & 10 PeV                & 30 PeV \\\hline
$f(100~m)$, part./m$^2$              & 0.0018            & 0.0048            & 0.0135                & 0.039 \\
S$_\mathrm{det}$, m$^2$             & 1667              & 625               & 222                   & 77                \\
J, part./(m$^2$$\cdot$s$\cdot$sr)    & $3.0\cdot10^{-6}$ & $2.6\cdot10^{-7}$ & $2.3\cdot10^{-8}$     & $2.0\cdot10^{-9}$ \\
N, event/year                       & 140000            & 12000             & 1050                  & 91                \\\hline
\end{tabular}
\end{table*}

It should be noted that the estimates of the number of registered events are lower estimates, since they assume that the particle density function does not depend on energy. In reality, with the growth of energy, the densities of secondary particles increase, and the distance from the shower axis at which 3 or more particles are registered within a fixed area of the setup also increases. This leads to a larger  effective area of EAS registration with higher energies.

\section{Joint registration of EAS and ultra-high energy muon}

This work does not include detailed modeling of the response of optical modules and trigger conditions of the Baikal-GVD~\cite{Avrorin:2020dre}. Instead, we use a simplified cluster trigger condition. We assume that the cluster is triggered if the energy released by a group of muons in the volume of water that the cluster occupies exceeds a certain threshold. We consider 3 different detection thresholds: 0.25, 0.5, and 1~TeV. The largest of these thresholds, 1~TeV, can be considered as a conservative estimate for the purpose of this work. Further studies are needed to refine the results, taking into account a complete modeling of the telescope response. 

To be registered by the telescope, muons need to travel a distance of at least 3.4~km, accounting for  the inclined geometry of simultaneous registration of muons and EAS. This gives us the estimate of the minimal muon energy ionization losses $\Delta E=3[\mathrm{MeV/cm}]\times3.4[\mathrm{km}] \approx$ 1~TeV. Taking these facts into account, we estimate the minimum energy of a muon on the surface that needs to be registered in the telescope to be of the order of 2~TeV.

\begin{figure}
\centering
\includegraphics[width=1.0\linewidth]{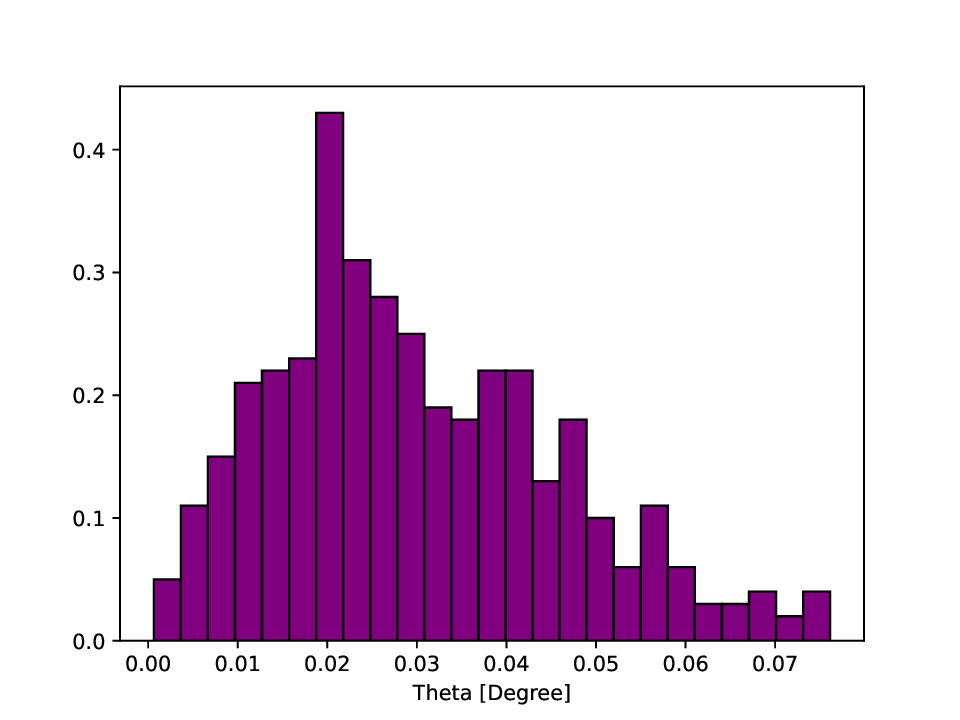}
\caption{\label{fig:Mugt2TeVangle} The distribution on angle difference between muon direction and EAS axis for muons with energy greater than 2~TeV in the 1~PeV EAS having zenith angle 76$^\circ$ .}
\end{figure}

In Figure~\ref{fig:Mugt2TeVangle}, we present the difference in angle between the direction of the muon and the EAS axis for muons with energy greater than 2~TeV in the 1~PeV EAS, having a zenith angle of 76$^\circ$ at ground level. According to simulations, the characteristic radius of the circle on the surface within which 50\% of these muons are contained is about 25~m. As might be expected, the bulk of such ultra-high energy muons propagate along the shower axis.

We propose the following idea of joint registration of a muon or muons in the GVD and the EAS that created them:
\begin{itemize}
    \item There must be events in the Baikal-GVD and onshore detection systems with a time difference that is consistent with the distance between the two systems and the probable angle of incidence of the EAS;
    \item The directions of the muons detected by the neutrino telescope and those detected by EAS should match;
    \item The energy of the EAS is determined by the amplitude of the signal in the onshore registration system and the distance to the shower axis, which is determined by using the direction of the registered muon(s) in the GVD. 
\end{itemize}

It should be noted that within our assumptions, the passage of a single muon with energy below the critical energy of 1~TeV cannot be registered in the Baikal-GVD cluster. The ionization losses of this particle are approximately 3~MeV/cm. Having a span length in the cluster of about 150~m gives us ionization losses of about 45~GeV. This is much less than the threshold energy of energy release in a single cluster used in this work (0.25, 0.5 or 1~TeV). for detection by a single cluster. For an event to be detected by Baikal-GVD, it must contain several muons, or the energy of a single muon must be much higher than the critical energy.

We investigated how the efficiency of atmospheric muon detection depends on the detection energy threshold in the Baikal-GVD cluster and on the energy of the initial EAS. For this purpose, for EAS-s with energies of 1, 3, 10, and 30 PeV we considered 3 different detection thresholds: 0.25, 0.5, and 1~TeV. The number of secondary muons in EAS with energies greater than 2~TeV and neutrinos with energies greater than 1~TeV, the Baikal-GVD detection efficiency for different thresholds in clusters, and the expected number of events per year of observations, together with an estimate of the detection area of the system, are presented in the Table~\ref{tab:EASMuDetdata}.

\begin{table*}
\caption{\label{tab:EASMuDetdata} Numbers of high energy muons and neutrinos in the EAS, detection efficiency and expected number of events for different EAS energies and detection thresholds in the GVD cluster.}
\centering
\begin{tabular}{l|c|c|c|c} \hline
E threshold               & 1 PeV   & 3 PeV     & 10 PeV    & 30 PeV \\\hline
S$_\mathrm{det}$, m$^2$   & 1667    & 625       & 222       & 77                \\
N$_\mu$(E$>$2~TeV)/EAS    & 3.9     & 9.2       & 21.6      & 59.6             \\
N$_\nu$(E$>$1~TeV)/EAS    & 3.5     & 8.7       & 20.9      & 57.3             \\
Efficiency ($\geq$1 cluster with $\Delta$E$>$0.25 TeV), \%  & 48     & 87       & 100      & 100             \\
Efficiency ($\geq$1 cluster with $\Delta$E$>$0.5 TeV), \%  & 27     & 63       & 98      & 100             \\
Efficiency ($\geq$1 cluster with $\Delta$E$>$1 TeV), \%  & 13     & 26       & 82      & 100             \\
N(EAS + $\geq$1 cluster with $\Delta$E$>$0.25 TeV), event/year  & 66000     & 10000       & 1050      & 91             \\
N(EAS + $\geq$1 cluster with $\Delta$E$>$0.5 TeV), event/year  & 37000     & 7500       & 1020      & 91             \\
N(EAS + $\geq$1 cluster with $\Delta$E$>$1 TeV), event/year   & 18000     & 3000       & 860      & 91             \\\hline
\end{tabular}
\end{table*}

\subsection{Joint registration of EAS and high energy neutrino}

We made estimations of the possibility to detect high energy atmospheric neutrinos together with the EAS that produced it:

\begin{itemize}
    \item number of EAS detected in a year with threshold energy $>$1~PeV -- N$_{\mathrm{EAS}} = 140000$ (Table~\ref{tab:EASMuDetdata}),
    \item number of neutrinos in the EAS (1~PeV) with E $>$ 1~TeV -- N$_\nu$ = 3.54 (Table~\ref{tab:EASMuDetdata}),
    \item cross section of the neutrino interaction at 1.5~TeV -- $\sigma = 1.5 \cdot 10^{-35}$ cm$^2$/nucleon \cite{Gandhi:1995tf},
    \item path length of neutrino detection in Baikal-GVD -- $l=1.35\cdot10^5$ cm,
    \item the probability that a neutrino will interact in the Baikal-GVD volume -- $\delta$N/N = n$\sigma l$ = $6 \cdot 10^{23} \times 1.5 \cdot 10^{-35} \times 1.35 \cdot 10^5 = 1.2 \cdot 10^{-6}$,
    \item finally, the number of expected simultaneously detected neutrinos -- $\Delta$N = N$_{\mathrm{EAS}}\cdot$N$_\nu\cdot \delta$N/N = $140000 \times  3.54 \times  1.2 \cdot  10^{-6}$ = 0.6 events per year.
\end{itemize}

As one can see, the suggested detector facility will not detect a reasonable number of EAS events together with an atmospheric neutrino in Baikal-GVD.

\section{Discussion}

According to the estimates presented, a ground-based detection system with an area of several hundred to thousands of square meters is required for the joint registration of high-energy muons and EAS. A complex with a detection area of thousands of square meters is comparable to modern experiments on the registration of EAS, and its creation requires an extensive list of original physical tasks and a serious budget.
A detector complex with an area of up to 600 square meters can be built in a reasonable time. This will allow the registration of up to 10000 inclined EAS per year, with energies from 3 PeV together with high-energy muons in Baikal-GVD. 

In our study, we did not address the issue of the type of detection system. Since the bulk of secondary particles in inclined EAS are muons, water Cherenkov detectors can be a good and cost-efficient choice. On the other side, scintillation counters have a higher efficiency as a result of their low detection threshold, which allows them to detect low-energy secondary electrons in the EAS. Scintillation counters could be tilted in such a way that their area is perpendicular to the main direction of the investigated inclined EASs. Tilted detectors will have an additional advantage, as the effective cross-section (area) for vertical events will be significantly smaller than that for inclined ones. Thus, events from vertical EASs will be effectively suppressed at the trigger level.

The actual choice of detection technology will depend on the requirements for the time resolution of the detectors and their size, due to the importance of these parameters for the trigger system.

\section{Conclusions}
Joint registration of an inclined EAS together with the high-energy muon produced in it is possible with full-scale Baikal-GVD and an onshore detector system having a registration area of 100--600~m$^2$. Depending on the size of the detector facility, the estimated number of registered joint events could range from 100 to 10000 events per year. The energy threshold for the registration of EAS will vary from 3 to 30 PeV. Such a setup would allow obtaining experimental information on the number of muons in EAS with ultra-high energy. This, subsequently, would make it possible to examine the hadronic interaction models used in EAS simulations and to verify the atmospheric neutrino flux calculations. Moreover, the setup will be able to simultaneously measure high- and low-energy muons in the same shower, which has been shown to be important for testing hadronic interaction models~\cite{Pyras:2025rjf}. The detector areas under consideration are insufficient for the joint registration of EAS and neutrinos and, as a consequence, for use as a veto detector. Another possible use of the detector is the cross-calibration of energy and direction measured by the Baikal-GVD.

\begin{acknowledgments}
This work is supported in the framework of the State project "Science"
by the Ministry of Science and Higher Education of the Russian Federation
under the contract 075-15-2024-54. The authors thank Dmitry Zaborov for fruitful consultation on the issues of particle registration features in Baikal-GVD and the features of modeling the interaction of high-energy muons with water.
\end{acknowledgments}

\end{document}